\documentclass[prb,twocolumn,aps,amsmath,amssymb]{revtex4}

\usepackage{graphicx}        % Include figure files

\begin{document}

\title{Difference between penetration and damping lengths in photonic crystal mirrors}

\author{C. Sauvan}
\email{christophe.sauvan@institutoptique.fr}
\author{J.P. Hugonin}
\author{P. Lalanne}
\affiliation{Laboratoire Charles Fabry de l'Institut d'Optique, CNRS, Univ. Paris-Sud, \\
Campus Polytechnique, RD128, 91127 Palaiseau, France}

\begin{abstract}
Different mirror geometries in two-dimensional photonic crystal slabs are studied with fully-vectorial calculations. We compare their optical properties and, in particular, we show that, for heterostructure mirrors, the penetration length associated with the delay induced by distributed reflection is not correlated to the characteristic damping length of the electromagnetic energy distribution in the mirror. This unexpected result evidences that the usual trade-off between short damping lengths and large penetration lengths that is classically encountered in distributed Bragg reflectors can be overcome with carefully designed photonic crystal structures.
\end{abstract}

\maketitle

Optical microcavities are able to trap light in small volumes close to the diffraction limit during a long time.~\cite{Nature-Vahala} They are essential components for various applications, including photonic integrated circuits~\cite{Nature-Lipson04} and semiconductor quantum light sources.~\cite{Nature-phot-Shields} Several recent experimental results have demonstrated that, among different light confinement schemes, Photonic Crystal (PhC) microcavities are highly valuable candidates for achieving extremely long lifetimes with wavelength-sized volumes.~\cite{Noda2,APL-Notomi,OE-Velha} The light confinement in these resonators can be understood with a simple Fabry-Perot model,~\cite{lpr-review} which highlights the key role played by PhC mirrors. Studying in details PhC-mirror properties is therefore essential for future applications and developments.

PhC mirrors are periodic structures in which the reflection of light is distributed over several periods. As such, they are usually thought to behave like classical Distributed Bragg Reflectors (DBRs), whose optical properties are well known and described in many textbooks.~\cite{Coldren-Corzine,Yariv-Yeh} This analogy is valid for most two-dimensional (2D) and one-dimensional (1D) PhC structures, but we evidence hereafter that some PhC mirrors may exhibit unusual properties that cannot be found in classical DBRs.

Let us consider the main mirror properties that are necessary to build a microcavity with a high quality factor and a small modal volume. The latter issue requires that the reflection be distributed over a short distance. More precisely, the evanescent tail of the electromagnetic energy distribution in the mirror has to be as small as possible. On the other hand, high quality factors are achieved provided that the reflection process be extremely efficient, i.e. with very low radiation losses.~\cite{lpr-review} However, reducing the losses is not the only mechanism for increasing the quality factor. Indeed, distributed reflection is not an instantaneous process. It induces a delay that can be used to increase the photon lifetime in the cavity. A mirror with a reflection coefficient $r=\sqrt{R}\exp(i\phi_r)$ reflects a light pulse with a delay $\tau = \partial\phi_r / \partial\omega$, which can also be expressed in terms of a characteristic length $L_p$. Usually called penetration length, $L_p$ is defined as the half-distance over which light would have propagated in the incident medium for accumulating the same delay,~\cite{Coldren-Corzine}

\begin{equation}\label{eq_lp}
L_p = \frac{1}{2}v_g\tau = \frac{c}{2n_g}\frac{\partial\phi_r}{\partial\omega} \,,
\end{equation}

\noindent where $v_g$ and $n_g$ are the group velocity and the group index of the incident mode. In classical DBRs, the penetration length defined by Eq.~(\ref{eq_lp}) is directly related to the characteristic damping length $\delta$ of the energy distribution inside the mirror. For arbitrary filling factors and index modulations $\Delta n$, the relation $L_p \leqslant \delta$ holds, the equality being fulfilled in the limit case of quarter-wave DBRs with low refractive index contrasts for which $L_p = \delta = n\Lambda/(2\Delta n)$, where $\Lambda$ is the grating period and $n$ is the average index.~\cite{Coldren-Corzine,Yariv-Yeh} Consequently, these two characteristic lengths are often considered to be equal and mixed-up in the literature. The delay induced by distributed reflection is therefore usually not seen as a degree of freedom to design high-Q and small-V microcavities, since large delays are thought to be synonymous of large volumes. However, we demonstrate hereafter that the penetration and damping lengths are not necessarily correlated and that carefully designed PhC mirrors allow to overcome the usual trade-off between short damping lengths and large penetration lengths.

\begin{figure}[t!]
 \centering
  \includegraphics[width= \columnwidth]{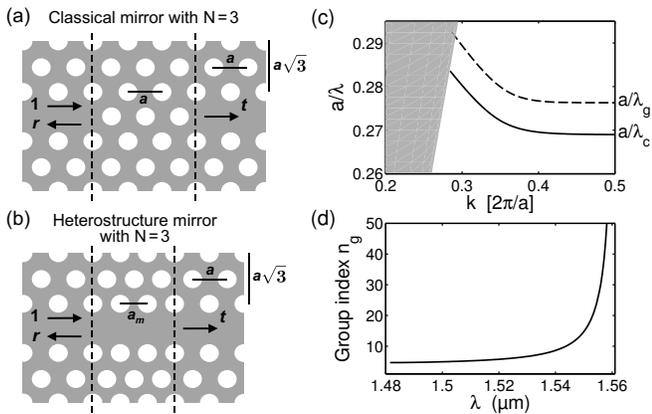}
  \caption{Schematic top view of the PhC mirrors and characteristics of the incident mode. The classical (a) and the heterostructure mirrors (b) are built in a triangular lattice of air holes (lattice constant $a = 420$ nm and hole radius $r = 126$ nm) etched in a silicon slab (refractive index $n = 3.42$ and thickness $h = 240$ nm). The longitudinal lattice constant in the heterostructure mirror is $a_m = 400$ nm. The number of periods in the mirrors is denoted by $N$. (c) Dispersion curves of the nominal W1 (solid curve) and the deformed W1 (dashed curve). Their cut-off wavelengths are $\lambda_c = 1.561~\mu$m and $\lambda_g = 1.52~\mu$m, respectively. The grey surface represents the range above the light line. (d) Group index of the incident mode (nominal W1).}
  \label{fig:schema}
\end{figure}

In this letter, we study the optical properties of reflectors in 2D PhC slabs with three-dimensional (3D) rigorous calculations. We compare different mirrors and demonstrate that, in the structure shown in Fig.~\ref{fig:schema}(b) and known in the literature as the heterostructure mirror,~\cite{Noda2} the penetration length associated with the delay is unexpectedly much larger than the characteristic damping length of the energy distribution. Thus, such mirrors can induce large delays while maintaining small volumes.

Figure~\ref{fig:schema} shows the two structures under study, which both aim at reflecting the fundamental mode of a single-line-defect PhC waveguide (W1 waveguide). In the first structure, the mirror consists of a finite-length section of the surrounding PhC, see Fig.~\ref{fig:schema}(a). It will be referred as the classical mirror hereafter. The second structure is the heterostructure mirror.~\cite{Noda2} It consists of a finite-length section of a deformed W1 waveguide with the same transverse lattice constant $a$, but a smaller longitudinal lattice constant $a_m < a$, see Fig.~\ref{fig:schema}(b). In both structures, the 2D PhC is a triangular lattice of air holes etched in a silicon membrane with $a = 420$ nm. The longitudinal lattice constant in the heterostructure mirror is $a_m = 400$ nm. The mirrors have a finite length $L = Na$ in Fig.~\ref{fig:schema}(a) and $L = Na_m$ in Fig.~\ref{fig:schema}(b), with $N$ the number of periods. Note that, in Fig.~\ref{fig:schema}(b), the waveguide is a nominal W1 and the mirror is a deformed W1. The incident guided modes are thus identical in Figs.~\ref{fig:schema}(a) and~\ref{fig:schema}(b), and we can directly compare the properties of both mirrors. Different behaviors can thus be unambiguously attributed to the mirrors themselves.

The main characteristics of the incident mode have been calculated below the light line and are summarized in Figs.~\ref{fig:schema}(c) and~\ref{fig:schema}(d). The computations have been performed with a 3D fully-vectorial Fourier Bloch-Mode Method (FBMM), which has been recently elaborated for analyzing light scattering in periodic waveguides.~\cite{OE-Guillaume} The dispersion curve of the incident mode is represented by the solid curve in Fig.~\ref{fig:schema}(c) and its group index is shown in Fig.~\ref{fig:schema}(d). The spectral range of interest is the spectral range of operation of the W1 below the light line. It extends from $\lambda = 1.48~\mu$m to $\lambda_c = 1.561~\mu$m, which corresponds to the incident mode cut-off at the edge of the first Brillouin zone, where the group index diverges. In the following, we are mainly interested in the optical properties of PhC mirrors in the region of moderate group index, $n_g < 20$. The classical mirror reflects light efficiently all over the spectral range of interest since this range naturally lies inside the band gap of the surrounding PhC. In the heterostructure, the mirror is a deformed W1 with a cut-off at $\lambda_g = 1.52~\mu$m. Its dispersion curve is represented by the dashed curve in Fig.~\ref{fig:schema}(c). The incident mode will be reflected only for wavelengths larger than the cut-off $\lambda_g$, where no propagative mode is available in the deformed W1. The cut-off $\lambda_g$ thus corresponds to the band-gap edge of the heterostructure mirror. The band gap width $\Delta\lambda = \lambda_c - \lambda_g$ decreases for decreasing lattice constant difference $\Delta a = a - a_m$.

\begin{figure}[t!]
  \includegraphics[width= \columnwidth]{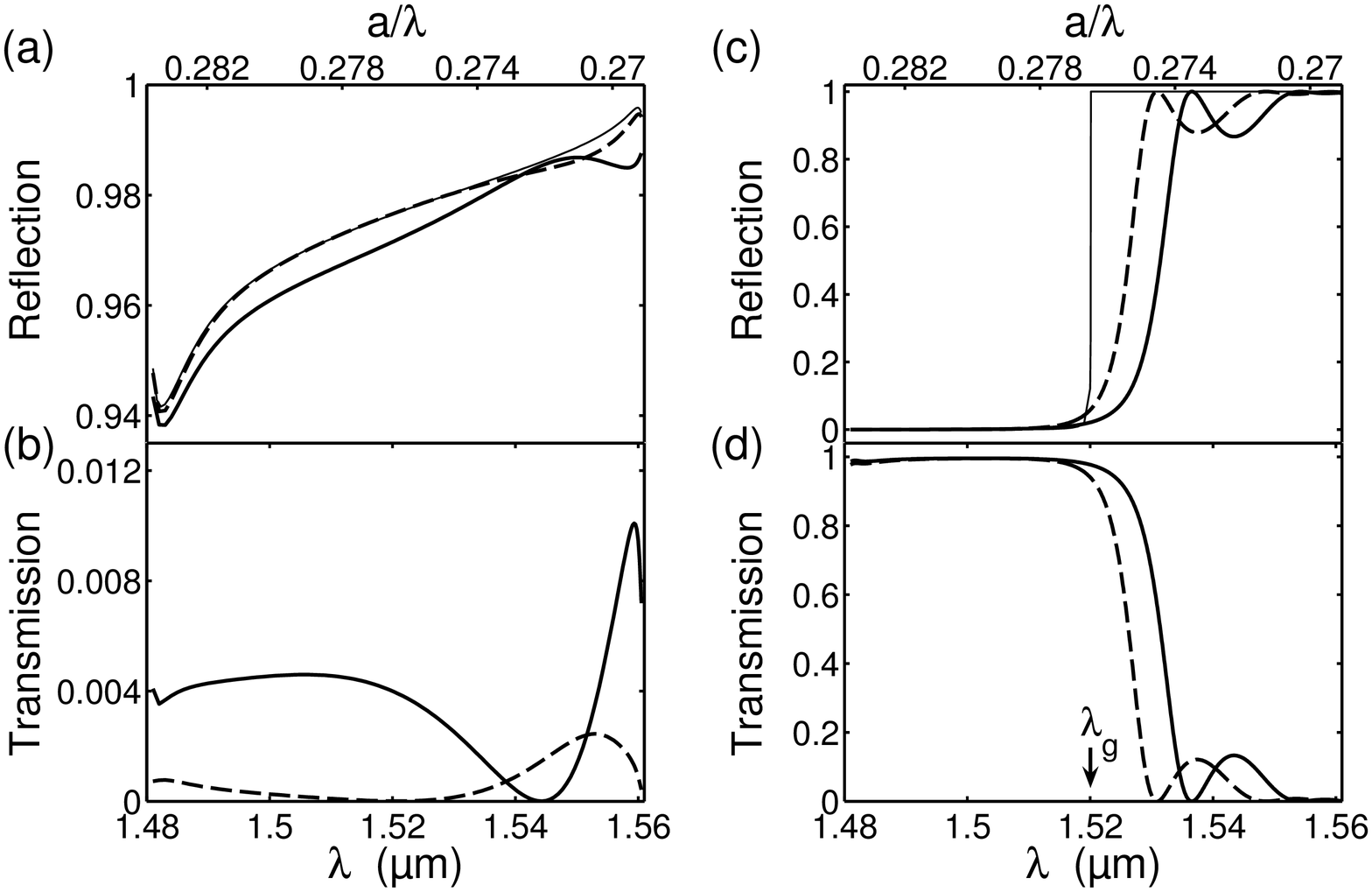}
  \caption{Reflection and transmission spectra. (a) Reflection and (b) transmission of the classical mirror. (c) Reflection and (d) transmission of the heterostructure mirror. The solid and dashed curves correspond to finite-length mirrors with $N = 3$ and $N = 4$, respectively. The thin curves in (a) and (c) correspond to semi-infinite mirrors.}
  \label{fig:RetT}
\end{figure}

The reflection and transmission coefficients of the PhC mirrors have been calculated with the FBMM.~\cite{OE-Guillaume} Figure~\ref{fig:RetT} shows their intensity $R = |r|^2$ and $T = |t|^2$ for two different lengths, $N = 3$ (solid curve) and $N = 4$ (dashed curve). The reflection and transmission of the classical mirror are represented in Figs.~\ref{fig:RetT}(a) and~\ref{fig:RetT}(b), and those of the heterostructure mirror are represented in Figs.~\ref{fig:RetT}(c) and~\ref{fig:RetT}(d). Heterostructure mirrors transmit all the incident light when operating outside their band gap, but their reflection increases rapidly as soon as $\lambda > \lambda_g$. The results evidence some oscillations in the spectra of finite-length mirrors. For some wavelengths, the transmission vanishes ($T \sim 0$) and the reflection reaches the value of the semi-infinite mirror (thin curves in Figs.~\ref{fig:RetT}(a) and~\ref{fig:RetT}(c)). The spectral position of the oscillations, which are much larger for the heterostructure, depends on the mirror length. Such a behavior inside the band gap does not exist in classical DBRs where the distributed reflection is due to a single evanescent Bloch mode.~\cite{Yariv-Yeh} But in the PhC mirrors under study, more than one evanescent Bloch mode are involved in the reflection process. These higher-order modes, which were previously reported for slow-light injection in W1 waveguides,~\cite{OL-Hugonin,OL-Australiens} cause some oscillations, whose amplitude decreases with the mirror length. Nevertheless, the radiation losses given by $1-R-T$ do not exhibit such oscillations; they are roughly equal for every length to $1-R_\infty$, with $R_\infty$ the reflection of the semi-infinite mirror (thin curve). It is noteworthy that heterostructure mirrors exhibit extremely low radiation losses,~\cite{Noda2,lpr-review} $1-R_\infty \sim 10^{-4}$.

\begin{figure}[t!]
  \includegraphics[width= \columnwidth]{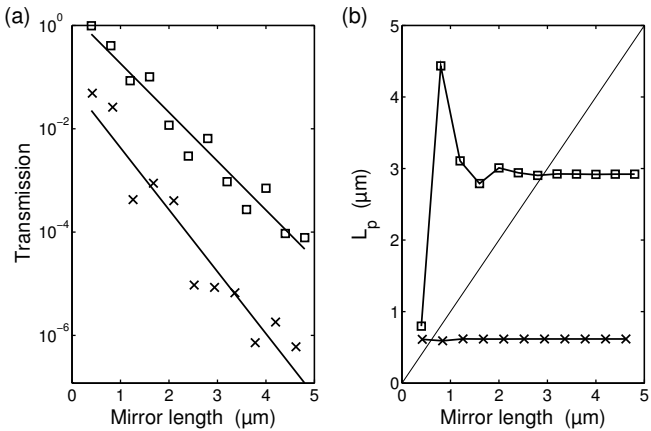}
  \caption{Mirror properties as a function of the mirror length for $\lambda = 1.54~\mu$m. (a) Transmission in a logarithmic scale. Crosses: classical mirror. Squares: heterostructure mirror. The solid lines show a linear fit of $\log(T)$ that allows to estimate the damping length $\delta$. (b) Penetration length. The thin line corresponds to $L_p = L$. The calculations have been performed from $N = 1$ to $N = 11$ periods.}
  \label{fig:TetLp_fN}
\end{figure}

Although several evanescent modes are involved in the reflection process, it is possible to estimate the characteristic damping length $\delta$ of the intensity in the mirror from the transmission calculation as a function of the mirror length. Figure~\ref{fig:TetLp_fN}(a) shows the variation of $T$ in a logarithmic scale for $\lambda = 1.54~\mu$m. The calculated data for the classical and the heterostructure mirrors are given by crosses and squares, respectively. As expected, the transmissions exponentially decay with the length, $T = \exp(-L/\delta)$,~\cite{Coldren-Corzine} and the linear fits shown by the solid lines in Fig.~\ref{fig:TetLp_fN}(a) allow to retrieve the damping length $\delta$ for both mirrors. They are of the same order of magnitude; $\delta = 360$ nm for the classical mirror and $\delta = 460$ nm for the heterostructure. The discrepancies between the ideal linear decrease and the actual data can be attributed to higher-order evanescent Bloch modes involved in the reflection process.

The penetration length $L_p$ is directly calculated from the phase of the reflected mode, see Eq.~(\ref{eq_lp}), and its variation with the mirror length for $\lambda = 1.54~\mu$m is represented in Fig.~\ref{fig:TetLp_fN}(b). For the classical mirror (crosses), $L_p = 620$ nm remains constant even for short mirrors ($N = 1$ or 2) and is comparable to the damping length. For the heterostructure mirror, the penetration length varies with the mirror length for very short mirrors, but it reaches rapidly a plateau corresponding to the value of the semi-infinite mirror. From $N = 3$, $L_p = 3~\mu$m is roughly constant and seven-fold larger than the damping length. Unlike the classical mirror that mostly behaves like an usual DBR, the heterostructure mirror exhibits an unexpected behavior with a penetration-length-to-damping-length ratio much larger than unity. Besides, it is interesting to note that short heterostructure mirrors may be remarkably efficient ($R > 99\%$) with penetration lengths that are even larger than the mirror length.

\begin{figure}[t!]
  \includegraphics[width= 0.95\columnwidth]{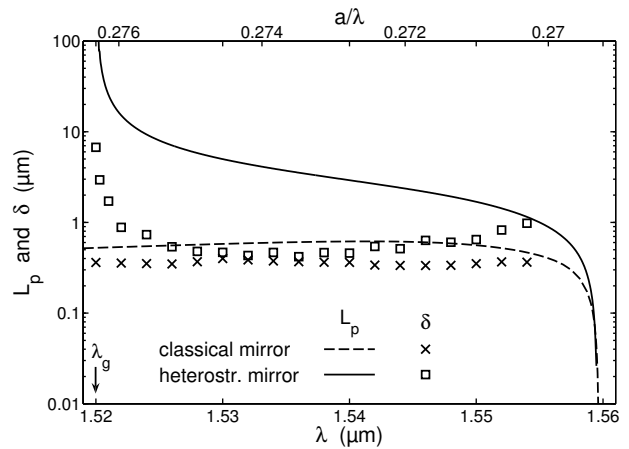}
  \caption{Spectral variation of the penetration length $L_p$ and of the damping length $\delta$. The calculations of the penetration length have been performed for semi-infinite mirrors. For the classical mirror, $L_p$ is represented by the dashed curve and $\delta$ by the crosses. For the heterostructure mirror, $L_p$ is represented by the solid curve and $\delta$ by the squares.}
  \label{fig:Lpetdelta}
\end{figure}

By repeating the fitting procedure shown in Fig.~\ref{fig:TetLp_fN}(a) for other frequencies, the spectral dependence of the damping length can be calculated. The results are represented in Fig.~\ref{fig:Lpetdelta} together with the penetration length. $L_p$ has been calculated for semi-infinite mirrors, since it weakly depends on the mirror length. For the classical mirror (dashed curve and crosses), both lengths are comparable, like in DBRs. On the contrary, for the heterostructure, the penetration length (solid curve) is much larger than the damping length (squares), except for large wavelengths where the incident group velocity is small. In fact, close to the band-gap edge $\lambda_g = 1.52~\mu$m, the properties of heterostructure mirrors are similar to those of classical DBRs, with a divergence of $\delta$ and $L_p$.~\cite{Yariv-Yeh} But the fact that $L_p$ remains one order of magnitude larger than $\delta$ inside the band gap is really unusual. This result evidences that, in a heterostructure mirror, light is reflected with a delay corresponding to a propagation distance that is considerably larger than the exponential decay of the intensity in the mirror.

In conclusion, we have studied the optical properties of reflectors in 2D PhC slabs. We have shown that, in heterostructure mirrors, the penetration length associated with the delay can be much larger than the characteristic damping length of the energy distribution inside the mirror. This amazing property evidences that these two characteristic lengths are not necessarily correlated. Consequently, the usual trade-off between short damping lengths and large penetration lengths that is encountered in classical DBRs can be overcome with carefully designed PhC structures. Intuitively, large penetration lengths may be understood as a result of the small lattice constant difference in the heterostructure. Indeed, such a behavior is classically encountered in DBRs with a small index contrast. But, contrary to DBRs, heterostructure mirrors exhibit also small damping lengths because of their peculiar dispersion properties.~\cite{NJP-Noda} This unexpected result allows to understand why, until now, the quality factors achieved in heterostructure cavities remain one order of magnitude larger than those achieved in 1D PhC resonators, while the mirror radiation losses and the cavity-mode volumes are essentially the same.~\cite{lpr-review}

%%%%%% Acknowledgments
The authors acknowledge funding through the EU-FP6 SPLASH project and the French ANR under contract Miraman ANR-06-NANO-000-30.

%%%%%% References


\begin{thebibliography}{13}
\expandafter\ifx\csname natexlab\endcsname\relax\def\natexlab#1{#1}\fi
\expandafter\ifx\csname bibnamefont\endcsname\relax
  \def\bibnamefont#1{#1}\fi
\expandafter\ifx\csname bibfnamefont\endcsname\relax
  \def\bibfnamefont#1{#1}\fi
\expandafter\ifx\csname citenamefont\endcsname\relax
  \def\citenamefont#1{#1}\fi
\expandafter\ifx\csname url\endcsname\relax
  \def\url#1{\texttt{#1}}\fi
\expandafter\ifx\csname urlprefix\endcsname\relax\def\urlprefix{URL }\fi
\providecommand{\bibinfo}[2]{#2}
\providecommand{\eprint}[2][]{\url{#2}}

\bibitem[{\citenamefont{Vahala}(2003)}]{Nature-Vahala}
\bibinfo{author}{\bibfnamefont{K.~J.} \bibnamefont{Vahala}},
  \bibinfo{journal}{Nature} \textbf{\bibinfo{volume}{424}},
  \bibinfo{pages}{839} (\bibinfo{year}{2003}).

\bibitem[{\citenamefont{Almeida et~al.}(2004)\citenamefont{Almeida, Barrios,
  Panepucci, and Lipson}}]{Nature-Lipson04}
\bibinfo{author}{\bibfnamefont{V.~R.} \bibnamefont{Almeida}},
  \bibinfo{author}{\bibfnamefont{C.~A.} \bibnamefont{Barrios}},
  \bibinfo{author}{\bibfnamefont{R.~R.} \bibnamefont{Panepucci}},
  \bibnamefont{and} \bibinfo{author}{\bibfnamefont{M.}~\bibnamefont{Lipson}},
  \bibinfo{journal}{Nature} \textbf{\bibinfo{volume}{431}},
  \bibinfo{pages}{1081} (\bibinfo{year}{2004}).

\bibitem[{\citenamefont{Shields}(2007)}]{Nature-phot-Shields}
\bibinfo{author}{\bibfnamefont{A.~J.} \bibnamefont{Shields}},
  \bibinfo{journal}{Nature Photonics} \textbf{\bibinfo{volume}{1}},
  \bibinfo{pages}{215} (\bibinfo{year}{2007}).

\bibitem[{\citenamefont{Song et~al.}(2005)\citenamefont{Song, Noda, Asano, and
  Akahane}}]{Noda2}
\bibinfo{author}{\bibfnamefont{B.~S.} \bibnamefont{Song}},
  \bibinfo{author}{\bibfnamefont{S.}~\bibnamefont{Noda}},
  \bibinfo{author}{\bibfnamefont{T.}~\bibnamefont{Asano}}, \bibnamefont{and}
  \bibinfo{author}{\bibfnamefont{Y.}~\bibnamefont{Akahane}},
  \bibinfo{journal}{Nature Materials} \textbf{\bibinfo{volume}{4}},
  \bibinfo{pages}{207} (\bibinfo{year}{2005}).

\bibitem[{\citenamefont{Kuramochi et~al.}(2006)\citenamefont{Kuramochi, Notomi,
  Mitsugi, Shinya, Tanabe, and Watanabe}}]{APL-Notomi}
\bibinfo{author}{\bibfnamefont{E.}~\bibnamefont{Kuramochi}},
  \bibinfo{author}{\bibfnamefont{M.}~\bibnamefont{Notomi}},
  \bibinfo{author}{\bibfnamefont{S.}~\bibnamefont{Mitsugi}},
  \bibinfo{author}{\bibfnamefont{A.}~\bibnamefont{Shinya}},
  \bibinfo{author}{\bibfnamefont{T.}~\bibnamefont{Tanabe}}, \bibnamefont{and}
  \bibinfo{author}{\bibfnamefont{T.}~\bibnamefont{Watanabe}},
  \bibinfo{journal}{Appl. Phys. Lett.} \textbf{\bibinfo{volume}{88}},
  \bibinfo{pages}{041112} (\bibinfo{year}{2006}).

\bibitem[{\citenamefont{Velha et~al.}(2007)\citenamefont{Velha, Picard,
  Charvolin, Hadji, Rodier, Lalanne, and Peyrade}}]{OE-Velha}
\bibinfo{author}{\bibfnamefont{P.}~\bibnamefont{Velha}},
  \bibinfo{author}{\bibfnamefont{E.}~\bibnamefont{Picard}},
  \bibinfo{author}{\bibfnamefont{T.}~\bibnamefont{Charvolin}},
  \bibinfo{author}{\bibfnamefont{E.}~\bibnamefont{Hadji}},
  \bibinfo{author}{\bibfnamefont{J.~C.} \bibnamefont{Rodier}},
  \bibinfo{author}{\bibfnamefont{P.}~\bibnamefont{Lalanne}}, \bibnamefont{and}
  \bibinfo{author}{\bibfnamefont{D.}~\bibnamefont{Peyrade}},
  \bibinfo{journal}{Opt. Express} \textbf{\bibinfo{volume}{15}},
  \bibinfo{pages}{16090} (\bibinfo{year}{2007}).

\bibitem[{\citenamefont{Lalanne et~al.}(2008)\citenamefont{Lalanne, Sauvan, and
  Hugonin}}]{lpr-review}
\bibinfo{author}{\bibfnamefont{P.}~\bibnamefont{Lalanne}},
  \bibinfo{author}{\bibfnamefont{C.}~\bibnamefont{Sauvan}}, \bibnamefont{and}
  \bibinfo{author}{\bibfnamefont{J.~P.} \bibnamefont{Hugonin}},
  \bibinfo{journal}{Laser and Photon. Rev.} \textbf{\bibinfo{volume}{2}},
  \bibinfo{pages}{514} (\bibinfo{year}{2008}).

\bibitem[{\citenamefont{Coldren and Corzine}(1995)}]{Coldren-Corzine}
\bibinfo{author}{\bibfnamefont{L.~A.} \bibnamefont{Coldren}} \bibnamefont{and}
  \bibinfo{author}{\bibfnamefont{S.~W.} \bibnamefont{Corzine}},
  \emph{\bibinfo{title}{Diode lasers and photonic integrated circuits}}
  (\bibinfo{publisher}{John Wiley}, \bibinfo{address}{New York},
  \bibinfo{year}{1995}).

\bibitem[{\citenamefont{Yariv and Yeh}(1984)}]{Yariv-Yeh}
\bibinfo{author}{\bibfnamefont{A.}~\bibnamefont{Yariv}} \bibnamefont{and}
  \bibinfo{author}{\bibfnamefont{P.}~\bibnamefont{Yeh}},
  \emph{\bibinfo{title}{Optical waves in crystals}} (\bibinfo{publisher}{John
  Wiley}, \bibinfo{address}{New York}, \bibinfo{year}{1984}).

\bibitem[{\citenamefont{Lecamp et~al.}(2007)\citenamefont{Lecamp, Hugonin, and
  Lalanne}}]{OE-Guillaume}
\bibinfo{author}{\bibfnamefont{G.}~\bibnamefont{Lecamp}},
  \bibinfo{author}{\bibfnamefont{J.~P.} \bibnamefont{Hugonin}},
  \bibnamefont{and} \bibinfo{author}{\bibfnamefont{P.}~\bibnamefont{Lalanne}},
  \bibinfo{journal}{Opt. Express} \textbf{\bibinfo{volume}{15}},
  \bibinfo{pages}{11042} (\bibinfo{year}{2007}).

\bibitem[{\citenamefont{Hugonin et~al.}(2007)\citenamefont{Hugonin, Lalanne,
  White, and Krauss}}]{OL-Hugonin}
\bibinfo{author}{\bibfnamefont{J.~P.} \bibnamefont{Hugonin}},
  \bibinfo{author}{\bibfnamefont{P.}~\bibnamefont{Lalanne}},
  \bibinfo{author}{\bibfnamefont{T.~P.} \bibnamefont{White}}, \bibnamefont{and}
  \bibinfo{author}{\bibfnamefont{T.~F.} \bibnamefont{Krauss}},
  \bibinfo{journal}{Opt. Lett.} \textbf{\bibinfo{volume}{32}},
  \bibinfo{pages}{2638} (\bibinfo{year}{2007}).

\bibitem[{\citenamefont{White et~al.}(2008)\citenamefont{White, Botten,
  de~Sterke, Dossou, and McPhedran}}]{OL-Australiens}
\bibinfo{author}{\bibfnamefont{T.~P.} \bibnamefont{White}},
  \bibinfo{author}{\bibfnamefont{L.~C.} \bibnamefont{Botten}},
  \bibinfo{author}{\bibfnamefont{C.~M.} \bibnamefont{de~Sterke}},
  \bibinfo{author}{\bibfnamefont{K.~B.} \bibnamefont{Dossou}},
  \bibnamefont{and} \bibinfo{author}{\bibfnamefont{R.~C.}
  \bibnamefont{McPhedran}}, \bibinfo{journal}{Opt. Lett.}
  \textbf{\bibinfo{volume}{33}}, \bibinfo{pages}{2644} (\bibinfo{year}{2008}).

\bibitem[{\citenamefont{Song et~al.}(2006)\citenamefont{Song, Asano, and
  Noda}}]{NJP-Noda}
\bibinfo{author}{\bibfnamefont{B.~S.} \bibnamefont{Song}},
  \bibinfo{author}{\bibfnamefont{T.}~\bibnamefont{Asano}}, \bibnamefont{and}
  \bibinfo{author}{\bibfnamefont{S.}~\bibnamefont{Noda}}, \bibinfo{journal}{New
  J. Phys.} \textbf{\bibinfo{volume}{8}}, \bibinfo{pages}{209}
  (\bibinfo{year}{2006}).

\end{thebibliography}
\end{document}